\documentclass{article}

\usepackage[final,sglblindworkshop]{neurips_2026}
\workshoptitle{AI for Chip Design}

\usepackage[utf8]{inputenc}
\usepackage[T1]{fontenc}
\usepackage{url}
\usepackage{booktabs}
\usepackage{graphicx}
\usepackage{amsmath}
\usepackage{amssymb}
\usepackage{nicefrac}
\usepackage{microtype}
\usepackage{xcolor}
\usepackage{hyperref}
\hypersetup{hidelinks}

\title{Independent Data-Integrity Verification for LLM-Driven Functional Coverage Closure in Hardware Design}

\author{%
  Venkata Krishna Kumar Vedantam \\
  Department of Electrical and Computer Engineering\\
  Portland State University\\
  \texttt{vedantam@pdx.edu} \\
}

\begin{document}

\maketitle

\begin{abstract}
LLM agents can write directed tests that close functional coverage, but coverage only shows that a state was reached, not that the data on the way there was correct. This is a long-known point in verification and we do not claim it as new. We describe a small coverage-closure agent that applies it to an LLM loop: coverage is read from the report that the simulation itself exports (cocotb-coverage YAML) and never taken from the model's account, and data integrity is checked by scoreboards that are separate from the model. On three architecturally distinct RTL designs (an asynchronous FIFO, a UART transmitter and an AHB2APB bridge) the agent raised coverage from 91.7\%, 88.9\% and 67.4\% to 100\% within its 8-iteration budget in every logged run, using one design-agnostic agent. A reactive scoreboard monitor attached to the FIFO's base testbench also found a clock-domain-crossing bug in the author's own stimulus code. The monitor is not yet attached to the LLM-generated tests, so their data integrity is not independently checked; one manual check on a single FIFO test is reported as preliminary. We list the weaknesses of this evidence: three designs, no baseline, no ablation, and logged runs that come from several agent checkpoints.
\end{abstract}

\section{Introduction}

Functional verification consumes roughly half of project time \citep{wilson2026}. The last uncovered bins, such as reset-timing corners and clock-domain-crossing (CDC) states, are hard to reach with random stimulus and often need hand-written directed tests. Recent work applies LLMs to this problem: testbench generation \citep{qiu2024autobench}, UVM component synthesis \citep{meng2026haven}, and coverage-gap repair loops \citep{uvm2_2026}. One benchmark of LLM stimulus generation from hardware specifications reports that frontier models produce stimulus that passes a golden RTL reference only 15.7--21.7\% of the time \citep{grpo_smu_2026}. That is a different task from ours, where the agent sees the RTL and gets simulator feedback, but it shows that LLM stimulus cannot be assumed correct.

That motivates the design choice in this paper: coverage (did a state occur) and data integrity (was the data correct) are two separate claims, and neither is taken from the LLM's own account of what it did.

\paragraph{Relation to prior work.} That coverage does not establish correct behavior is well known. For RTL code coverage, observability-based metrics were proposed for this reason: a statement counts as covered only if an error there would propagate to an observable output \citep{devadas1996occom,fallah1998occom}. Our bins are functional coverage, not code coverage, so that work is related by analogy and not directly applicable, and our agent uses no such metric. It keeps coverage and checking as separate measurements and uses scoreboards for checking, which is standard practice; the contribution is applying that separation to an LLM test-generation loop and being explicit about where it does and does not hold. HAVEN \citep{meng2026haven} avoids free-form LLM code by expanding an extracted specification with a template engine. UVM2 \citep{uvm2_2026} uses a coverage-gap-driven repair loop. Here the LLM writes cocotb test code directly. We have not compared against these systems.

\section{Method}

\begin{figure}[t]
  \centering
  \includegraphics[width=\linewidth]{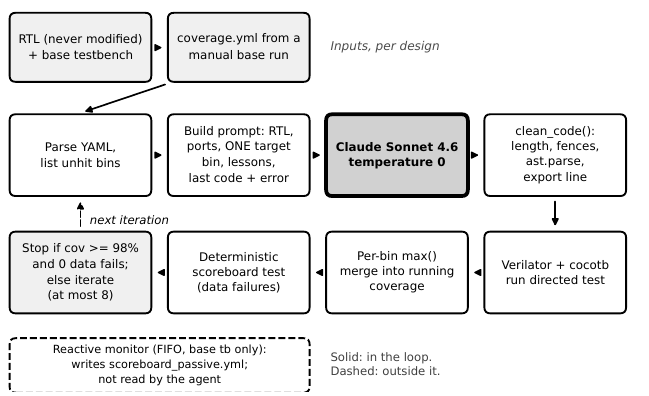}
  \caption{The agent. An engineer runs the base testbench once to produce \texttt{coverage.yml}; the agent then loops until the threshold is met with zero data failures or 8 iterations have run. The reactive monitor is outside the loop.}
  \label{fig:arch}
\end{figure}

\paragraph{Inputs.} Each design has an RTL directory, a base cocotb testbench, and a short config file. The base testbenches are written once per design by the author and are not generated by the agent. The agent requires a \texttt{coverage.yml} (cocotb-coverage YAML) from a manual run of the base testbench and stops if it is missing. The agent code contains no design-specific signal or bin names.

\paragraph{Loop (Figure~\ref{fig:arch}).} (1) Parse \texttt{coverage.yml} and list the unhit bins. (2) Pick one of them as the target for this iteration, cycling through the list by iteration index. (3) Build a prompt with the RTL source, the port list extracted from the RTL to reduce invented signal names, the named target bin, up to 12 unhit bins, the last five lessons from a persistent \texttt{LESSONS.md}, and the previous test code with its simulation error. (4) Ask the model for a directed cocotb test. (5) Run \texttt{clean\_code()}: length check, markdown-fence removal, \texttt{ast.parse}, removal of any coverage-export line at module level (it would write empty coverage at import time), and addition of an export call inside the test if none is present. (6) Run the test with Verilator. After every successful simulation, merge the new coverage into a running result with a per-bin \texttt{max()}, so a bin once hit is never lost; this is the union-merge idea of UCIS \citep{ucis2012}. (7) Run a separate deterministic scoreboard test, and stop when coverage is at least the threshold (98\%) with zero data failures, or after 8 iterations. If the budget runs out, the agent stops and reports the remaining unhit bins for an engineer; it does not escalate automatically.

\paragraph{Single-bin target.} Each request carries one explicit objective, a named uncovered bin, instead of an open-ended request to cover everything that is left. This was added in checkpoint v2.5 because a prompt listing all unhit bins spread the model's effort; we saw better test generation informally during development, but did not measure or ablate the choice. The target is a soft instruction in the prompt: the model's test may hit other bins, and the merge credits them.

\paragraph{Two checkers.} The \emph{deterministic scoreboard test} runs fixed, known stimulus and records which bins pass or fail a data check; the data-failure count that the stop rule uses comes from it. It does not run the LLM's stimulus. A zero there means the design's data path behaved correctly for known inputs, and says nothing about whether the LLM-generated tests are correct. The \emph{reactive monitor} watches DUT signals in concurrent coroutines during whatever stimulus is actually driven. It exists only for the FIFO, is attached only to the base testbench, and writes a file the agent never reads.

\paragraph{Model and execution.} Claude Sonnet 4.6 (\texttt{claude-sonnet-4-6}) through the Anthropic API directly, temperature 0, \texttt{max\_tokens} 8000. There is no coding harness or tool use; each iteration is one request. The agent prints token counts per call but does not store them. For one final-version FIFO run, a slow one that needed 7 model calls, the counts were roughly 64k input and 21k output tokens, about \$0.5 at list prices (\$3 and \$15 per million tokens for Sonnet 4.6). This is a single run, not an average.

\section{Results}

\begin{table}[t]
\caption{Closure results. Bins are functional-coverage bins. Every logged run ended at 100\% within the 8-iteration budget. Runs that stalled with a stale \texttt{LESSONS.md} are described in the text and are not in the table.}
\label{tab:main}
\centering
\small
\begin{tabular}{lrccc}
\toprule
Design & Bins & Base (hit/total) & Final & Fault tests \\
\midrule
Async FIFO & 36 & 91.7\% (33/36) & 100\% & 5/5 \\
UART TX    & 18 & 88.9\% (16/18) & 100\% & 4/4 \\
AHB2APB    & 46 & 67.4\% (31/46) & 100\% & 5/5 \\
\bottomrule
\end{tabular}
\end{table}

\paragraph{Closure.} The results come from 12 logged closure runs with a stated iteration count that we could trace in the checkpoint notes (v2.9 to v2.10.2), run on two machines (Apple Silicon and GitHub Codespaces on x86\_64). A line repeated verbatim in a later checkpoint is counted once. The earlier v2.3 run predates single-bin targeting and lesson memory and is left out. Every one of these runs ended at 100\% within the 8-iteration budget; 10 of the 12 needed 4 iterations or fewer, and the slowest needed 8 (FIFO and AHB2APB). Counts are as written in the run notes and can differ by one from the number of LLM calls. These are not repeated trials of one fixed agent. According to the checkpoint notes the agent loop has not changed since v2.10 apart from a shorter sleep, while the scoreboards, base testbench and monitor changed in later checkpoints up to the released v2.10.3.1. The agent closed the 3, 2 and 15 bins that the FIFO, UART and AHB2APB base testbenches left uncovered (8\%, 11\% and 33\% of their bins).

Runs started with a \texttt{LESSONS.md} left over from an earlier session stalled at the base coverage (91.7\% on the FIFO). We attribute this to the stale entries, did not isolate the cause, and did not record how many such runs occurred. We clear the file before every run, so the closure results above are for runs started with a cleared file, and the file acts only as memory within one run.

\paragraph{Scoreboards.} Data failures were 0 in the runs recorded with the corrected scoreboards (v2.10.2 and later); earlier runs predate those fixes and count only toward closure. As a sanity check that the scoreboards can fail, we wrote 14 scoreboard self-tests by fault injection (FIFO 5, UART 4, AHB2APB 5). In each, the DUT is driven correctly and the scoreboard is told to expect a corrupted value: a bit flip, zeros, off-by-one, inversion, reordering, a phantom write, or a read before the synchronization wait. All 14 were flagged. The fault classes were chosen by hand, not by any statistical argument, so this shows only that the scoreboards are not vacuous. It is not mutation testing: the RTL is never changed.

\paragraph{The CDC bug.} The base FIFO testbench drove the read enable (\texttt{r\_en}) from a loop timed by the write clock (10~ns) in three of its phases, while the FIFO's read logic runs on an independent read clock (15~ns). The DUT then occasionally performed extra reads and produced real data mismatches. The RTL was not modified; the bug was in the author's testbench code, and the reactive monitor found it, not the LLM. The testbench had been used for more than 50 days of development before the monitor existed; this is calendar time, and we did not count how many tests had run. The fix drives reads from dedicated read-clock coroutines. After the fix, successive verification rounds on 12, 100 and 1000 random seeds showed zero FIFO scoreboard failures.

\paragraph{Monitor on LLM-generated stimulus.} As a one-time manual check, we attached the monitor to one LLM-generated directed FIFO test, the one closing the three hardest CDC bins, and 463 of 463 checks passed. This was a single test checked by hand, to see whether the monitor works on directed-style stimulus. It is not part of the automated loop and does not show that LLM-generated tests are generally correct.

\section{Limitations and Future Work}

\paragraph{What is established.} On three small designs, the agent closed the coverage the base testbenches missed, using coverage read from the report the simulation exports. The scoreboards flag the 14 injected faults. The reactive monitor found one real stimulus bug.

\paragraph{What is not.} (i) The LLM-generated directed tests are not independently checked in the loop. Only one test was checked by hand with the monitor. A test can close coverage with invalid stimulus, and we do not measure how often that happens. (ii) There are three small designs and 12 logged runs from several agent checkpoints, with no fixed-version repeated trials, no confidence intervals and no pass@$k$. (iii) There is no baseline, such as plain random re-seeding, a scripted strategy, or another LLM system, and no ablation of the single-bin target, \texttt{LESSONS.md} or \texttt{clean\_code()}. These three were each added during development (the export-line fix in \texttt{clean\_code()} after Codespaces runs wrote empty coverage files), not chosen by experiment, and we did not measure how much each helps. (iv) Only one model was used. (v) The base testbenches are hand-built per design, so the agent's input is not free. Base coverage therefore describes our testbenches, not a measured random-stimulus plateau; the 1000-seed runs mentioned above checked data integrity, not coverage. (vi) If \texttt{clean\_code()} rejects the output, for example on a syntax error, the iteration is skipped and the model is not told why.

\paragraph{Next steps.} Attach the monitor to every generated test and report how often a coverage-closing test is rejected on data integrity; this also requires checking the monitor for false alarms on reset-heavy tests. Run repeated trials of one fixed version with a random-seed baseline, and ablate the single-bin target. Make the lesson memory specific to the target bin instead of global. Add multiple model families. A bounded repair step for rejected tests is planned and not implemented.

\paragraph{LLM usage disclosure.} An LLM (Claude, Anthropic) is the core, non-incidental subject and mechanism of this work: it generates the directed test stimulus described throughout Section 2. All coverage and data-integrity claims attributed to the LLM's output are independently verified against real simulation output as described; none are taken from the model's own self-report.

\paragraph{Code.} \url{https://github.com/KrishnaKumarVedantam/AMBU-CovAgent/}

{\small
\bibliographystyle{plainnat}
\bibliography{references}
}

\end{document}